# A theoretical prediction on huge hole and electron mobilities of 6,6,18-graphdiyne nanoribbons

Hongyu Ge, Guo Wang[*] and Yi Liao

*Department of Chemistry, Capital Normal University, Beijing 100048, China*

[*]E-mail address: wangguo@mail.cnu.edu.cn

ABSTRACT

Two-dimensional 6,6,18-graphdiyne and the corresponding one-dimensional nanoribbons are investigated using crystal orbital method. Based on HSE06 functional, the one-dimensional confinement increases the band gaps. With band gaps larger than 0.4 eV, thirty-three 6,6,18-graphdiyne nanoribbons have larger majority carrier mobilities at room temperature than the highest value of armchair graphene nanoribbons. Unlike γ-graphdiyne, 6,6,18-graphdiyne nanoribbons have both huge hole and electron mobilities, depending on whether they are armchair or zigzag type. The huge mobilities are explained by crystal orbital analysis. The superior capabilities of 6,6,18-graphdiyne nanoribbons make them possible candidates for high speed electronic devices in complementary circuits.

*Keywords:* 6,6,18-graphdiyne nanoribbon; carrier mobility; band gap; crystal orbital; HSE06 functional.

## 1. Introduction

Graphdiyne [1], mostly known as γ-graphdiyne, has attracted extensive attention [2] since it was synthesized [3]. Graphdiyne is a new carbon allotrope beyond graphene, which is one of the reasons for its success. And more important, it has many predicted appealing properties, such as huge electron mobilities [4,5], high capabilities in gas purification and lithium batteries [2].

In electronic industry, carrier mobility is an important issue, which directly affects the speed of electronic devices. Current complementary circuits technology in computer industry takes the advantage of low power, but requires both large hole and electron mobilities of electronic devices. Compared with silicon, armchair graphene nanoribbons (AGNR) are predicted to have much larger hole or electron mobilities, depending on their width [6,7]. Therefore, AGNR could be good candidates for next-generation electronic devices. However, atomically precise controlling the width of the nanoribbons in top-down methods is still a challenge. The bottom-up synthetic method makes graphdiyne another carbon-based candidate for high speed electronic devices [3].

A finite band gap, preferably larger than 0.4 eV [8] is essential to achieve high on/off ratio in electronic devices. Although two-dimensional γ-graphdiyne and the corresponding one-dimensional nanoribbons have band gaps larger than 0.4 eV, the predicted electron mobilities are always larger than the hole mobilities [4,5]. It is desirable to find a material with both large hole and electron mobilities, for ease of fabrication in electronic industry. Previous works focused mainly on γ-graphdiyne [4,5,9-28] and a few focused on α-graphdiyne [29-31]. However, there should be other types of graphdiynes that also have two acetylenic linkages (diyne). The linkages in graphdiyne introduce rich geometrical variety and should make graphdiynes promising materials for various potential applications.

In this work, two-dimensional 6,6,18-graphdiyne (denoted as GDY18) as well as its one-dimensional nanoribbons are constructed and investigated using crystal orbital method based on density functional theory. It is indicated that the nanoribbons with band gaps larger than 0.4 eV have robust huge hole and electron mobilities at room temperature, which are explained with crystal orbitals.

## 2. Models and computational details

Two-dimensional GDY18, which has six, eighteen and twenty-two-membered rings, is shown in Figure 1. Two-dimensional γ-graphdiyne was synthesized via a cross-coupling reaction using hexaethynylbenzene [3]. For GDY18, the building blocks of can be viewed as hexaethynylbenzene and tetraethynylethene indicated by rectangles in Figure 1. For the one-dimensional counterpart, following the convention of graphene nanoribbons, both armchair and zigzag GDY18 nanoribbons (denoted as

AGDY18NR and ZGDY18NR) are constructed. Considering the building blocks, the nanoribbons are always ended with ethynyl groups. The number $N$ in Figure 1 represents the number of building blocks.

Since a sufficiently large band gap is essential to achieve high on/off ratio in operation of an electronic device, accurately describing the band gaps is important in this work. The screened hybrid functional HSE06, which can give precise band gaps for solids [32], is used throughout the work. Bloch functions based on a standard 6-21G($d$, $p$) basis set in CRYSTAL14 program [33,34] are used. Full geometrical optimizations of the structures are performed before calculating the properties. A large atomic grid with 75 radial and 974 angular points is used in density functional numerical integration. Five parameters 8, 8, 8, 8, 18 are used to control the high accuracy of bielectronic integrals. Dense Monkhorst-Pack samplings with 81 k-points in the first Brillouin zones are used. The precision is sufficiently to obtain the converged geometries and related properties of the structures. Especially, the k-point samplings are ten times denser when calculating the band structures, in order to facilitate the fitting of carrier effective masses. It is noted that when there is not a sharp density of states near frontier band edges, carriers in a range wider than $k_\mathrm{B}T$ should participate in the conduction. It this work, the energy range 10 $k_\mathrm{B}T$ [35] is used to fit the effective masses. It should be noted that calculations based on plane wave basis [36] also give similar geometries and band structures, and there is no spin effect on the structures.

Under the deformation potential theory for semiconductors [37], when wavelength of an electron is much larger than a lattice constant, carriers are mostly scattered by longitudinal acoustic phonons with long wavelengths. For the GDY18NR, this assumption is still valid and will be discussed below. The carrier mobilities of the one-dimensional GDY18NR are obtained by [38]

$$\mu_\mathrm{1D} = \frac{e\hbar^2 C_\mathrm{1D}}{(2\pi k_\mathrm{B}T)^{1/2}|m^*|^{3/2} E_1^2}, \tag{1}$$

where $C_\mathrm{1D}$ is a stretching modulus for a one-dimensional crystal, $m^*$ is a carrier effective mass and $E_1$ is a deformation potential constant. Carrier effective masses are obtained by fitting frontier bands, while stretching moduli and deformation potential constants are obtained under deformed geometries [39]. The deformation potential theory is successfully applied to similar systems, such as carbon nanotube [40,41] and graphene [6,7,42].

## 3. Results and discussions

The calculated energy of the two-dimensional GDY18 per atom is only 2 meV higher than that of γ-graphdiyne, implying its considerable stability and high possibility of being synthesized. The band structures in Figure 2(a) indicate that GDY18 is a semiconductor with a narrow band gap at X' point. The band gap (0.33 eV) is too small [8] to achieve high on/off ratio. Quantum confinement in one-dimensional nanoribbons with finite width may increase its band gap and make it suitable for electronic devices.

Direct band gaps occur for all the nanoribbons. Since the band structures of all the AGDY18NR or ZGDY18NR are similar, only 6-AGDY18NR and 6-ZGDY18NR are taken as examples. As shown in Figure 2(b) and 2(c), the valence band maximum (VBM) and conduction band minimum (CBM) are at Γ point of the first Brillouin zone for all the AGDY18NR, while they are at X point for all the ZGDY18NR. This is quite different from γ-graphdiyne nanoribbons, which always have direct band gaps at Γ point for both armchair and zigzag patterns [4,5].

The band gaps of 2-AGDY18NR and 2-ZGDY18NR are as large as 1.75 and 2.12 eV, respectively. The narrow nanoribbons have much larger band gaps than the two-dimensional GDY18. As shown in Figure 3(a), the band gaps decrease monotonically with the width for both armchair and zigzag nanoribbons. The nanoribbons 13-AGDY18NR and 27-ZGDY18NR are the widest structures with band gaps larger than 0.4 eV. Their gaps are 0.41 and 0.40 eV as indicated by arrows in Figure 3(a). The band gaps decrease further with the width and should reach 0.33 eV for the two-dimensional GDY18.

The room temperature carrier mobilities are calculated under the deformation potential theory for the nanoribbons with band gaps larger than 0.4 eV. The carrier velocities near the frontier band edges for AGDY18NR are in the range of 0.71-2.05×10$^5$ ms$^{-1}$. The corresponding de Broglie wavelengths are in the range of 35-102 Å, which are much longer than the lattice constants (9.44-9.46 Å). The situation for ZGDY18NR is similar. The carrier velocities are in the range of 0.50-1.39 ×10$^5$ ms$^{-1}$. The corresponding wavelengths are in the range of 52-145 Å, also much longer than the lattice constants (13.82-13.88 Å). These verify that the deformation potential theory can be used to calculate the carrier mobilities of the nanoribbons.

The stretching moduli increase almost linearly with the width. The slopes are 8.9 and 12.8 eVÅ$^{-2}$ for $N$-AGDY18NR ($N$=2-13) and $N$-ZGDY18NR ($N$=2-27), respectively, indicating higher mechanical strength for the zigzag nanoribbons. These values are much smaller than that of the armchair graphene nanoribbons (25.0 eVÅ$^{-2}$). Because of the $sp$ hybridized linkages, the arrangement of carbon atoms in GDY18 is sparser than in graphene. The density of GDY18 is only 56% magnitude as large as that of graphene. The latter is very light with a density of 0.76 mg per m$^2$. Since the carrier mobility depends only linearly on the stretching modulus as shown in equation (1) and the difference of the stretching modulus is not large, the influence coming from the stretching modulus should be limited.

A deformation potential constant is obtained by an energy change at VBM or CBM with respect to lattice deformation proportion. As shown in Figure 3(b), all the valence band deformation potential constant $E_{1v}$ (0.78-1.18 eV) are smaller than the conduction band deformation potential constant $E_{1c}$ (4.57-4.70 eV) for AGDY18NR. Unlike the stretching moduli, the deformation potential constants significantly affect the carrier mobilities as indicated in equation (1). The very small $E_{1v}$ should produce very large hole mobilities and make AGDY18NR favourable to hole transport. For ZGDY18NR, all the $E_{1v}$ (3.01-5.60 eV) are larger than $E_{1c}$ (0.37-1.32 eV), making them suitable for electron transport.

Crystal orbitals can explain the reason for the unbalanced deformation potential constants. In Figure 4(a) and 4(b), the orientations of the highest occupied crystal orbital (HOCO) at VBM or the lowest unoccupied crystal orbital (LUCO) at CBM of 6-AGDY18NR are different, especially on some carbon atoms with bonds along the one-dimensional periodic direction. As indicated by rectangles in Figure 4(a), the HOCO on some atoms are parallel to the one-dimensional direction. However, vertical component exists in the LUCO on the same atoms. The delocalized orbitals in the HOCO should have smaller energy change than the localized ones in the LUCO during the deformation along the periodic direction [6,7], so the $E_{1v}$ is smaller than the $E_{1c}$. As for 6-ZGDY18NR, the situation is just opposite. The HOCO on some carbon atoms in Figure 4(c) are localized as indicated by rectangles, while the LUCO in Figure 4(d) is delocalized. Therefore, the $E_{1v}$ is larger than the $E_{1c}$.

The hole and electron effective masses are similar for the same AGDY18NR or ZGDY18NR. Since AGDY18NR (ZGDY18NR) should be favourable to hole (electron) transport, only hole (electron) effective masses for AGDY18NR (ZGDY18NR) are shown in Figure 3(c). The hole effective mass of 2-AGDY18NR is 0.65 $m_0$. The mass fast decreases to 0.24 $m_0$ for 4-AGDY18NR and gradually decreases to about 0.14 $m_0$ for the wide nanoribbons. The electron effective mass of 2-ZGDY18NR is as big as 3.70 $m_0$. Then it fast decreases to 0.30 $m_0$ for 4-AGDY18NR and finally decreases slowly to 0.13 $m_0$ for the wide nanoribbons. The heavy carriers for the narrow nanoribbons should be due mainly to the quantum confinement. When the nanoribbons are wider, the confinement becomes weaker and the carriers move more freely. The light carriers in the wide nanoribbons should produce large carrier mobilities.

The majority carrier mobilities (hole mobility $\mu_h$ for AGDY18NR and electron mobility $\mu_e$ for ZGDY18NR) are shown in Figure 3(d). The carrier mobilities increase with the width, because of the larger stretching moduli and the smaller carrier effective masses for the wider nanoribbons. For AGDY18NR and ZGDY18NR with the same number of building blocks $N$, the mobilities are comparable. When $N$=2-13, the $\mu_h$ for AGDY18NR increase from $1.4 \times 10^3$ to $2.0 \times 10^5$ cm$^2$V$^{-1}$s$^{-1}$, while the $\mu_e$ for ZGDY18NR increase from $1.1 \times 10^2$ to $2.9 \times 10^5$ cm$^2$V$^{-1}$s$^{-1}$. For ZGDY18NR with $N$=14-27, the $\mu_e$ increase further to $1.8 \times 10^6$ cm$^2$V$^{-1}$s$^{-1}$. Unlike γ-graphdiyne nanoribbons that are only favourable to electron transport, GDY18NR have both huge hole and electron mobilities. This should be useful for next-generation carbon-based high speed complementary circuits.

The GDY18NR and AGNR are compared in order to investigate the reason for the huge carrier mobilities. GDY18NR have *sp* hybridized linkages, so the connection between carbon atoms is not as dense as that in AGNR with all *sp*$^2$ hybridized carbon atoms. The number of bonds in GDY18NR is also smaller, so the energy change of the orbitals should be smaller than that in AGNR. Furthermore, many *sp* hybridized linkages shown in Figure 4 have no frontier crystal orbital distributed on. All these make the deformation potential constants extremely small, especially for the majority carriers (0.78-1.18 eV for holes of AGDY18NR and 0.37-1.32 eV for electrons of ZGDY18NR). For AGNR with denser atoms, these values are about 2-5 eV calculated with the same method.

The majority carrier mobilities of AGNR are also increase with the width. The widest AGNR with a band gap larger than 0.4 eV is 34-AGNR. It has an $E_{1v}$ of 4.14 eV and the majority carrier (hole) mobility is $1.8 \times 10^4$ cm$^2$V$^{-1}$s$^{-1}$. Although AGNR have larger stretching moduli, the influence from the deformation potential constants is more significant according to equation (1). The majority carrier mobilities of the widest 13-AGDY18NR and 27-ZGDY18NR with band gaps larger than 0.4 eV are $2.0 \times 10^5$ and $1.8 \times 10^6$ cm$^2$V$^{-1}$s$^{-1}$. These values are one or two order larger than that of 34-AGNR, indicating that GDY18NR are superior to AGNR in carrier mobilities.

In fact, the majority carrier mobilities of 4-AGDY18NR ($1.9 \times 10^4$) and 5-ZGDY18NR ($2.2 \times 10^4$) are already larger than that of 34-AGNR. The huge hole and electron mobilities of GDY18NR are robust. With the band gaps larger than 0.4 eV, there are thirty-three GDY18NR that have carrier mobilities larger than the upper bound of AGNR. They are 4-AGDY18NR to 13-AGDY18NR with width of 26.6-90.2 Å and 5-ZGDY18NR to 27-ZGDY18NR with width of 27.0-130.8 Å. For AGNR, the width determines the carrier types [6]. Considering that all the AGDY18NR are favourable to hole transport while ZGDY18NR are favourable to electron transport, exactly controlling the width is not necessary for GDY18NR to adjust carrier types.

## 4. Conclusions

Two-dimensional GDY18 and the corresponding one-dimensional AGDY18NR and ZGDY18NR are constructed and investigated using crystal orbital method based on density functional theory. The energy of GDY18 per atom is only 2 meV higher than that of γ-graphdiyne, indicating its considerable stability. Its band gap is too small to achieve high on/off ratio for electronic devices. Quantum confinement in the one-dimensional nanoribbons increases their band gaps. There are thirty-eight nanoribbons (2-AGDY18 to 13-ZGDY18 and 2-ZGDY18 to 27-ZGDY18 ) with band gaps larger than 0.4 eV. Direct band gaps exist at Γ or X point of the first Brillouin zone for all the AGDY18NR or ZGDY18NR, respectively. This is quite different from armchair and zigzag γ-graphdiyne nanoribbons, which always have direct band gaps at Γ point. The deformation potential constants are also different. For AGDY18NR, the $E_{1v}$ is smaller than the $E_{1c}$. While for ZGDY18NR, the situation is opposite. These make AGDY18NR and ZGDY18NR favourable to hole and electron transport, respectively. GDY18NR have both high hole and electron mobilities, depending on their orientation. This is also different from γ-graphdiyne nanoribbons that always have high electron mobilities. Therefore, GDY18NR are superior to γ-graphdiyne nanoribbons. The majority carrier mobilities of 13-AGDY18NR and 27-ZGDY18NR with band gaps larger than 0.4 eV are $2.0 \times 10^5$ and $1.8 \times 10^6$ cm$^2$V$^{-1}$s$^{-1}$, one or two order larger than the highest value for AGNR. Therefore, GDY18NR are superior to AGNR in carrier mobilities. Unlike AGNR, only orientation controls the carrier types of GDY18NR. There are ten AGDY18NR and twenty-three ZGDY18NR with the carrier mobilities larger than that of AGNR. Both robust huge hole and electron mobilities indicate that GDY18NR are possible candidates for high speed electronic devices in complementary circuits.

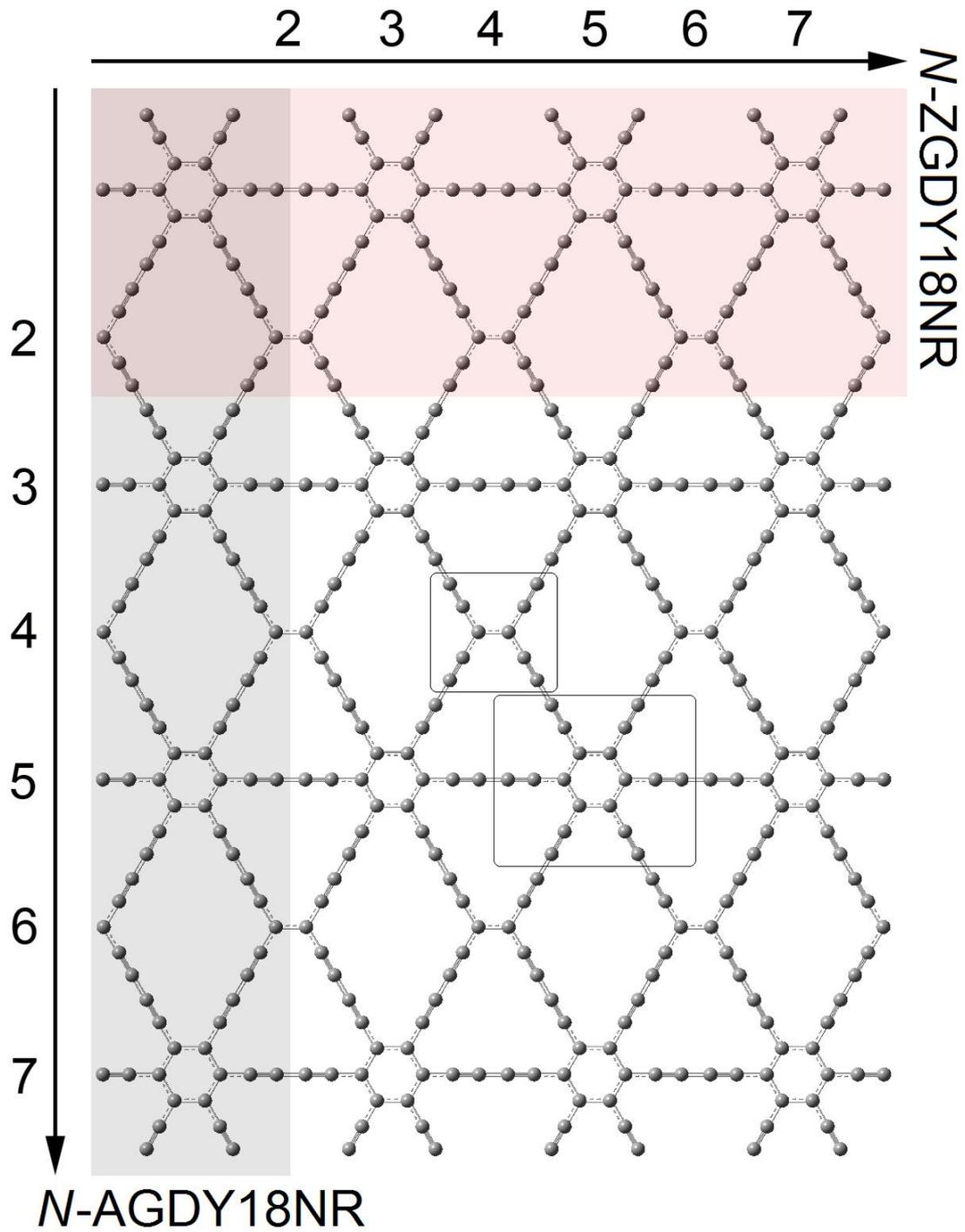

**Figure 1.** Model of two-dimensional GDY18. The unit cells of *N*-AGDY18NR and *N*-ZGDY18NR are presented in shadow.

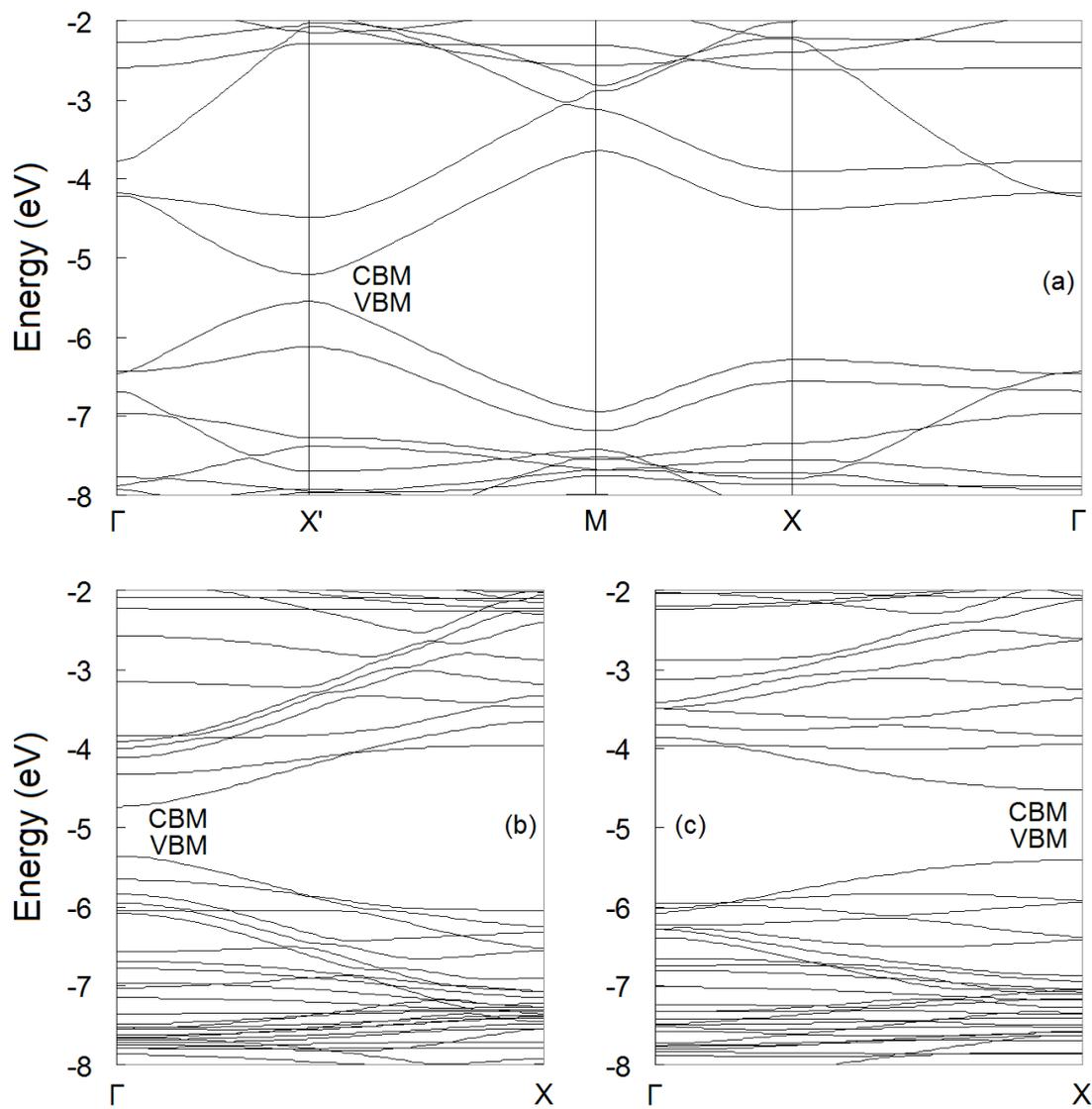

**Figure 2.** Band structures of (a) two-dimensional GDY18, (b) 6-AGDY18NR and (c) 6-ZGDY18NR.

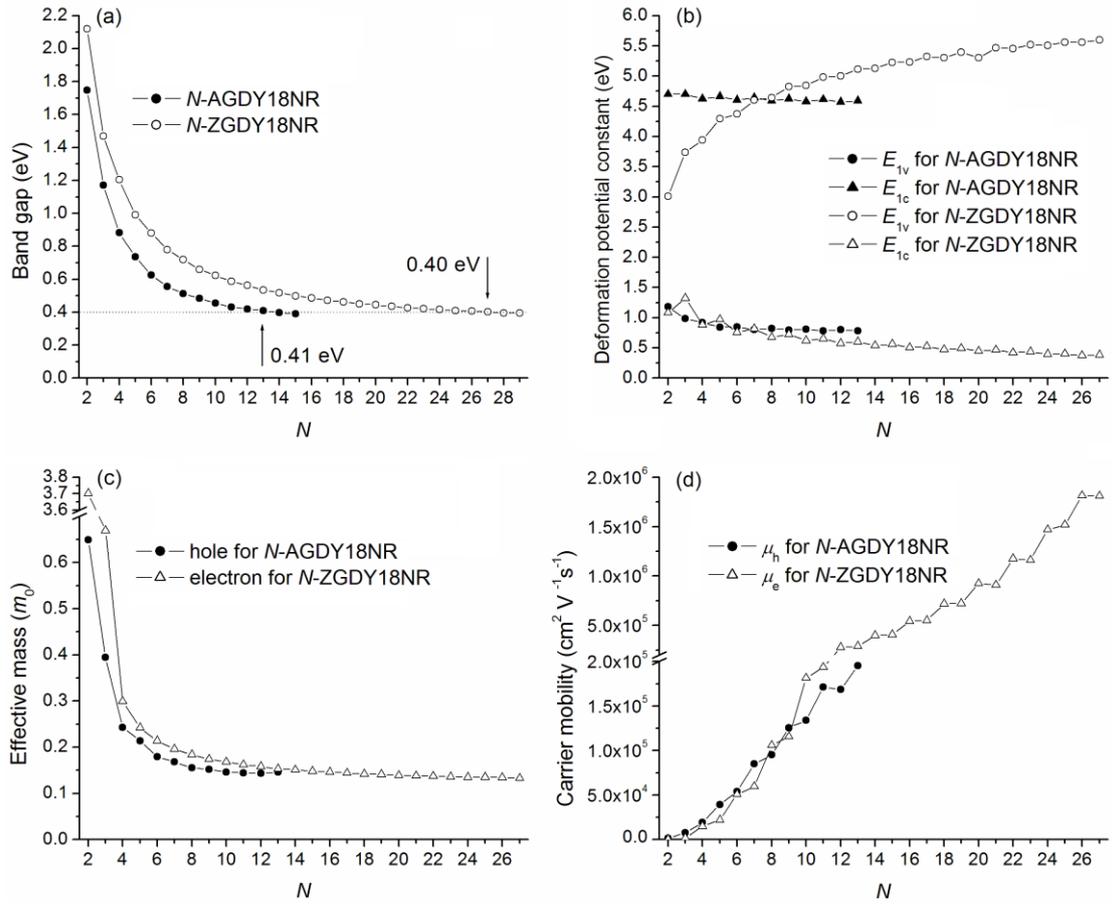

**Figure 3.** (a) Band gaps, (b) deformation potential constants, (c) effective masses and (d) carrier mobilities of GDY18NR.

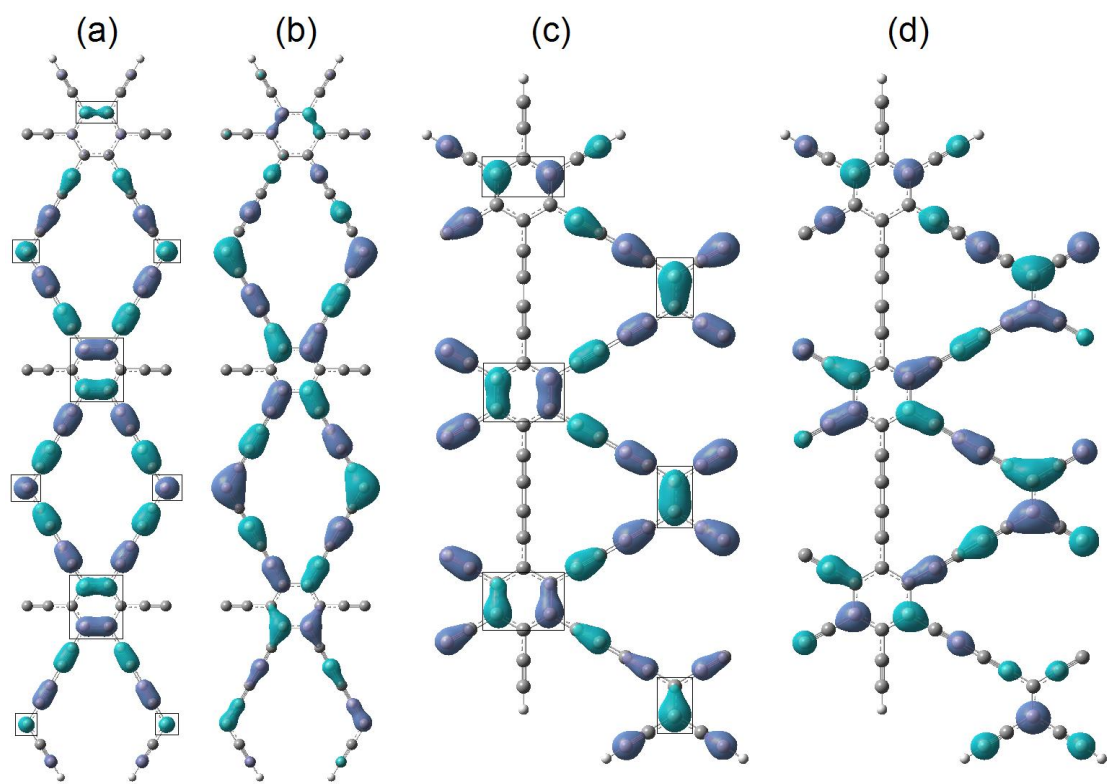

**Figure 4.** (a) HOCO and (b) LUCO of 6-AGDY18NR, (c) HOCO and (d) LUCO of 6-ZGDY18NR.